\documentclass[11pt,a4paper]{article}
\usepackage[hyperref]{acl2018}
\usepackage{times}
\usepackage{latexsym}

\usepackage{url}
\usepackage{graphicx}

\aclfinalcopy 

\title{ASR-based Features for Emotion Recognition: A Transfer Learning Approach}

\author{No\'e Tits, Kevin El Haddad, Thierry Dutoit \\
  Numediart Institute, \\
  University of Mons, Belgium \\
  {\tt \{noe.tits, kevin.elhaddad, thierry.dutoit\}@umons.ac.be} \\}

\date{}

\begin{document}
\maketitle
\begin{abstract}


During the last decade, the applications of signal processing have drastically improved with deep learning. However areas of affecting computing such as emotional speech synthesis or emotion recognition from spoken language remains challenging.
In this paper, we investigate the use of a neural Automatic Speech Recognition (ASR) as a feature extractor for emotion recognition.  We show that these features outperform the eGeMAPS feature set to predict the valence and arousal emotional dimensions, which means that the audio-to-text mapping learned by the ASR system contains information related to the emotional dimensions in spontaneous speech. We also examine the relationship between first layers (closer to speech) and last layers (closer to text) of the ASR and valence/arousal.
\end{abstract}

\section{Introduction}

With the advent of deep learning, areas of signal processing have been drasctically improved. In the field of speech synthesis, Wavenet \cite{wavenet-16-vandenoord}, a deep neural network for generating raw audio
waveforms, outperforms all previous approaches in terms of naturalness.
One of the remaining challenges in speech synthesis is to control its emotional dimension (happiness, sadness, amusement, etc.). The work described here is part of a larger project to control as accurately as possible, the emotional state of a sentence being synthesized. For this, we present here exploratory work regarding the analysis of the relationship between the emotional states and the modalities used to express them in speech.

Indeed one of the main problems to develop such a system is the amount of good quality data (naturalistic emotional speech of synthesis quality, i.e. containing no noise of any sorts). This is why we are considering solutions such as synthesis by analysis and transfer learning \cite{transfer_learning-2010-pan}.

Arousal and valence \cite{circumplex-80-russel} are among the most, if not the most used dimensions for quantizing emotions. Valence represents the positivity of the emotion whereas arousal represents its activation. Since they represent emotional states, these dimensions are linked to several modalities that we use to express emotions (audio, text, facial expressions, etc.).

It has recently been shown that for emotion recognition, deep learning based systems learn features that outperform handcrafted features \cite{adieu_features-16-trigeorgis} \cite{affect_modeling-13-martinez} \cite{speech_emotion_recognition_skip_connections-17-kim,emotion_recognition_in_the_wild-17-kim}.
The use of context and different modalities has also been studied with deep learning models. \citet{context_dependent-17-poria} focus on the contextual information among utterances in a video while \citet{tensor_fusion-17-zadeh, memory_fusion-18-zadeh} develop specific architectures to fuse information coming from different modalities.

In this work, with the goal to study the relationship between valence/arousal, and different modalities, we propose to use the internal representation of a speech-to-text system. An Automatic Speech Recognition (ASR) system or speech-to-text system, learns a mapping between two modalities: an audio speech signal and its corresponding transcription. We hypothesize that such a system must also be learning representations of emotional expressions since these are contained intrinsically in both speech (variation or the pitch, the energy, etc.) and text (semantic of the words).

In fact, we show here that the activations of certain neurons in an ASR system, are useful to estimate the arousal and valence dimensions of an audio speech signal. In other words, transfer learning is leveraged by using features learned for an automatic speech recognition (ASR) task to estimate valence and arousal. The advantage of our method is that it allows combining the use of large datasets of speech with transcriptions with limited datasets annotated in emotional dimensions.

An example of transfer learning is the work of \citet{openai-17-radford}. They trained a multiplicative LTSM \cite{mLSTM-16-krause} to predict next character based on the previous ones to design a text generator system. The dataset used to train their model was the Amazon review dataset presented in \citet{amazon_reviews-15-mcauley}. Then, they used the representation learned by the model to predict sentiment also available in the dataset, and achieved state of the art prediction.
 
In this paper, we show that the activations of a deep learning-based ASR system trained on a large database can be used as features for the estimation of arousal and valence values. The features would therefore be extracted from both the audio and text modalities which the ASR system learned to map.




\section{ASR-based Features for Emotion Prediction Via Regression}

Our goal is to study the relationship between valence/arousal, and audio/text modalities thanks to an ASR system. The main idea is that the ASR system that models the mapping between audio and text might learn a representation of emotional expression. So, for our analyses, we use an ASR system as a feature extractor which feeds a linear regression algorithm to estimate the arousal/valence values.
This section describes the whole system. First we present the ASR system used as a feature extractor. We then briefly present the data used and present first results on the data analysis.

\subsection{ASR system}

The ASR system used is implemented in \cite{stt_wavenet-16-namju} and pre-trained on the VCTK dataset \cite{vctk-17-veaux} containing 44 hours of speech uttered by 109 native speakers of English.  

Its architecture consists of a dilated convolution of blocks. Each block is a gated constitutional unit (GCU) with a skip (residual) connection. In other words a Wavenet-like architecture \cite{wavenet-16-vandenoord}. There are 15 layers and 128 GCUs in each layer: 1920 GCUs in total.

To lighten the computational cost, the audio signal is compressed in 20 Mel-Frequency Cepstral Coefficients (MFCCs) and then fed into the system.

\subsection{Dataset Used}
\subsubsection*{IEMOCAP Dataset}
The "interactive emotional dyadic motion capture database" (IEMOCAP) dataset \cite{iemocap-08-busso} is used in this paper. It consists of audio-visual recordings of 5 sessions of dialogues between male and female subjects. In total it contains 10 speakers and a total of 12 hours of data. The data is segmented in utterances. Each utterance is transcribed and annotated by category of emotions \cite{basic_emotions-92-Ekman} and a value for emotional dimensions \cite{circumplex-80-russel} (valence, arousal and dominance) between 1 and 5 representing the dimension's intensity.

In this work, we only use the audio and text modalities as well as the valence and arousal annotations.


\subsubsection*{Data Analysis and Neural Features}

\begin{figure}[h]
\centering
\includegraphics[scale=0.65]{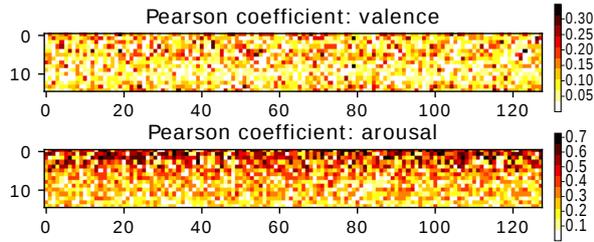}
\caption{Pearson correlation coefficient between the neural features and valence (up) and arousal (down) - Female speaker of session 2}
\label{corr_va_neurons_S2F}
\end{figure}

We investigate the relationship between the activation output of the ASR-based system's GCUs and the valence/arousal values by studying the correlations between them. For every utterance and for each speaker of the IEMOCAP dataset, we compute the mean activation of the GCUs of the ASR. The Pearson correlation coefficient is then calculated between the mean activation outputs and the values of valence/arousal of all utterances of the speaker. In the rest of the paper, we will refer to the mean activation of the GCUs as neural features.
As an example, the results concerning the female speaker of  session 2 is summarized in a heat map represented in Figure~\ref{corr_va_neurons_S2F}

Each row of the heat map corresponds to a layer of GCUs. The color is mapped with the Pearson correlation coefficient value.

One can see that correlations exist for both arousal and valence. This suggests that the ASR-based system learns a certain representation of the emotional dimensions.

\subsection{Structure of the system}

\begin{figure}[h]
\centering
\includegraphics[scale=0.2, width = 0.5\textwidth]{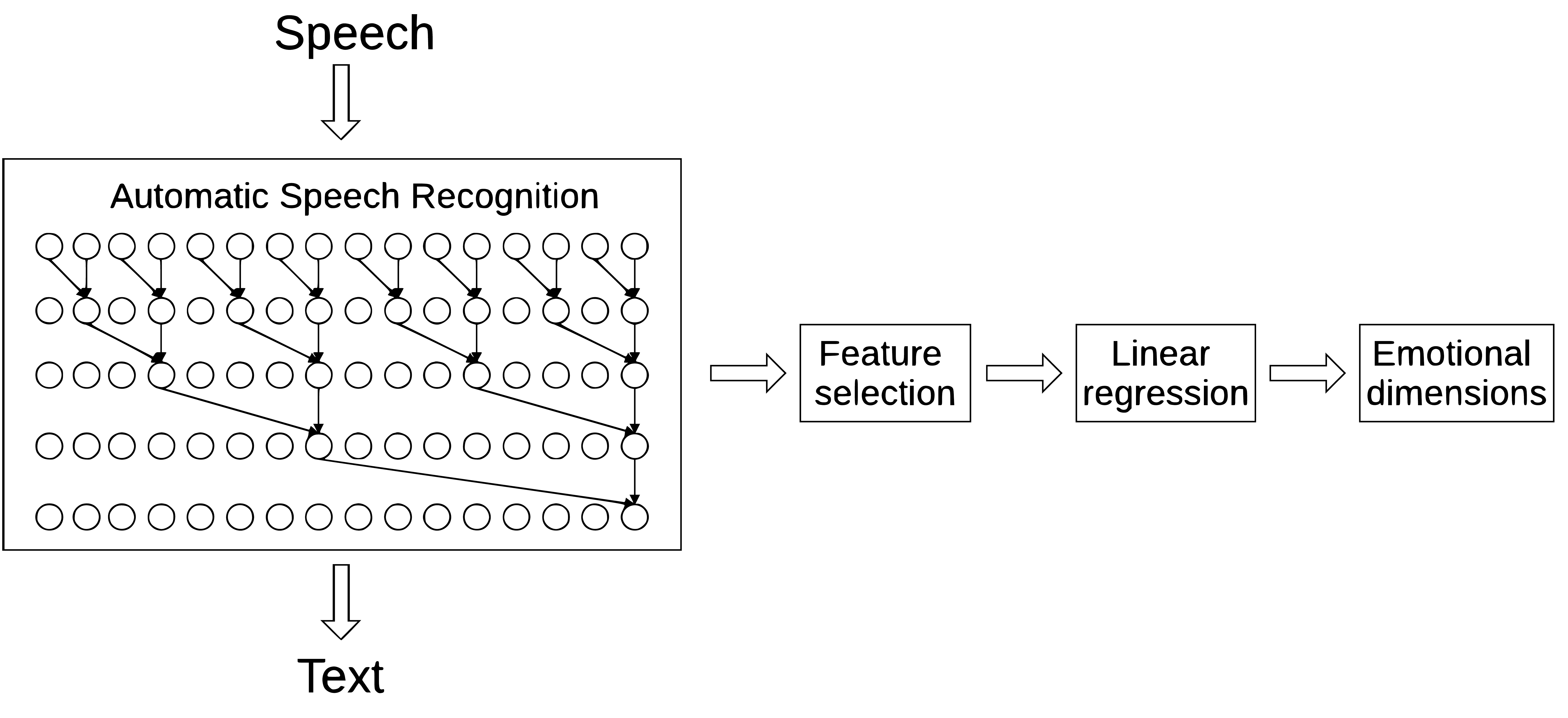}
\caption{Block diagram of the system}
\label{system}
\end{figure}

The system is illustrated in Figure~\ref{system}. As previously mentioned, the ASR system is used as a feature extractor. First we compute the 20 MFCCs of the utterances of the IEMOCAP dataset with librosa python library \cite{librosa-15-mcfee}. These are passed through the ASR to compute the corresponding neural features.

A feature selection is applied on the neural features to keep 100 among the 1920 features for dimensionality reduction purpose. The selection is done using the scikit-learn python library  \cite{scikit-learn-11-pedregosa} with the Fisher score. 

Finally a linear regression is trained to estimate the valence/arousal values from the neural features using the IEMOCAP data. The linear regression is done using scikit-learn. The training is done by minimizing the Mean Squared Error (MSE) between predictions and labels.

\section{Experiments and Results}

In this section, we detail the experiments that we carried out. The first one is the evaluation of the neural features in terms of MSE and its comparison with a linear regression of the eGeMAPS feature set \cite{egemaps-16-eyben}. In the second one, we investigate the relationship between the audio and text and modalities and the emotional dimensions.

\subsection{First experiment: Linear regression}

In this first experiment,  we investigate the performance of a linear regression to predict arousal and valence using the neural features. We compare this with a linear regression using the eGeMAPS feature set.

The eGeMAPS feature set is a selection of acoustic features that provide a common baseline for evaluation in researches to avoid differences of feature set and implementations. Indeed, they also provide their implementation with openSMILE toolkit \cite{opensmile-10-eyben} that we used in this work.

The features were selected based  on  their ability to represent affective  physiological  nuances  in  voice  production, their proven performance in former research work as well as the possibility to extract them automatically, and their theoretical significance.

The result of this selection is a set of 18 Low-level descriptors (LLDs) related to frequency (pitch, formants etc.), energy (loudness, Harmonics-to-Noise Ratio, etc.) and spectral balance (spectral slopes, ratios between formant energies, etc.). Then several functionals such as standard deviation and mean are applied to these LLDs to have the final features.

The results obtained from the linear regression in terms of MSE are compared to the annotations for each of the arousal and valence values (between 1 and 5) in Table~\ref{MSE}.

\begin{table}[h]
\centering
\small
\begin{tabular}{|c|c|c|c|c|}
\hline
&\multicolumn{2}{|c|}{Arousal} & \multicolumn{2}{|c|}{Valence} \\
\hline
& Mean & Variance & Mean & Variance\\
\hline
Neural features	& \textbf{0.259}& 	0.020&	\textbf{0.660}&	0.118\\
eGeMAPS set&	0.267&	0.034&	0.697&	0.135\\
\hline
\end{tabular}
\caption{MSE on the prediction of valence and arousal.}\label{MSE}
\end{table}

We perform a leave-one-speaker-out evaluation scheme with both feature sets for cross-validation. In other words, each validation set in constituted with the utterances corresponding to one speaker and the corresponding training set with the other speakers. We train a model with each training set and evaluate it on the validation set in terms of MSE. The table contains the mean and standard deviation of the MSEs.

It is clear from this table that the neural features outperform the eGeMAPS in this experiment. This confirms the fact that the ASR system learns representations of emotional dimensions in spontaneous speech.


\subsection{Second experiment: Influence of modalities}

\begin{figure}[h]
\centering
\includegraphics[scale=0.65]{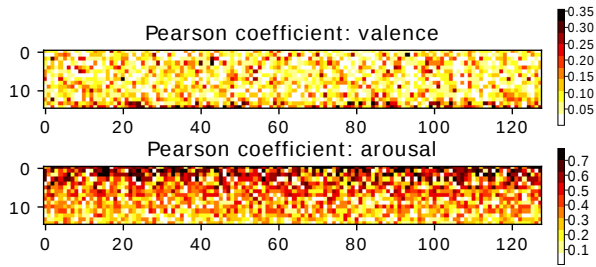}
\caption{Pearson correlation coefficient between the neural features and valence (up) and arousal (down) - Female speaker of session 1}
\label{corr_va_neurons_S1F}
\end{figure}

During the data exploration, we noticed that, for some speakers, the layers closer to the speech input were more correlated to arousal and the ones closer to the text output to valence. An example is shown in Figure~\ref{corr_va_neurons_S1F}. We present, in this section, preliminary studies regarding this matter.

In order to analyze this phenomenon as precisely as possible, we only considered the utterances from the IEMOCAP database for which the valence/arousal annotators were consistent with each other, leaving us with 7532 utterances in total instead of 10039.

Then we performed linear regression with 4 different sets of feature to study their influence. For the first set, we select the 100 best features among the 3 first layers of the neural ASR in terms of Fisher score using scikit-learn. For the second set, we apply the same selection to the 3 last layers. The third set selection is applied among all neural features. The last set is the eGeMAPS feature set.


The results are summarized in Figure~\ref{MSE_layers}. As expected, the results show, that for the speakers considered, the layers closer to the audio modality outperform the ones closer to the text modality in the ASR architecture for arousal prediction and vice versa for the valence prediction. On this we build a hypothesis that the arousal-related features learned are more related to the audio modality than the text and vice versa for the valence-related features. This hypothesis will be further explored in future work.

\begin{table}[h]
\centering
\small
\begin{tabular}{|c|c|c|c|c|}
\hline
&\multicolumn{2}{|c|}{Arousal} & \multicolumn{2}{|c|}{Valence} \\
\hline
& Mean & Variance & Mean & Variance\\
\hline

First layers &	0.325&	0.069&	0.714&	0.114\\
Last layers &	0.357&	0.038&	0.661&	0.089\\
All &	0.296&	0.044&	0.621&	0.099\\
eGeMAPS set&	0.328&	0.064&	0.683&	0.124\\

\hline
\end{tabular}
\caption{Means and variances of the MSE on the prediction of valence and arousal.}\label{MSE_layers}
\end{table}

\section{Conclusions and Future work}

In this paper, we show that features learned by a deep learning-based system trained for the Automatic Speech Recognition task can be used for emotion recognition and outperform the eGeMAPS feature set, the state of the art handcrafted features for emotion recognition. Then we investigate the correlation of the emotional dimensions arousal and valence with the modalities of audio and text of the speech. We show that for some speakers, arousal is more correlated to neural features extracted from layers closer to the speech modality and valence to the ones closer to the text modality.

To improve the system, we plan to perform an end-to-end training including the average operation. Another avenue to explore is to replace the average over time by a max-pooling over time which according to \citet{max_pool_time-17-aldeneh} select the frames that are emotionally salient.

Then an analysis of the underlying activation evolutions could be done to see if it is possible to extract a frame-by-frame description of valence and arousal without having to annotate a database frame-by-frame.

Concerning the second experiment, we intend to investigate why these correlation patterns are only visible for some speakers and not others and the relationship between the arousal/valence and audio/text.
We thereby hope to better understand the way multidimensional representations of emotions can be used to control the expressiveness in synthesized speech.





\section*{Acknowledgments}


No\'e Tits is funded through a PhD grant from the Fonds pour la Formation \`a la Recherche dans l'Industrie et l'Agriculture (FRIA), Belgium. 


\newpage
\bibliography{acl2018}
\bibliographystyle{acl_natbib}

\end{document}